\documentclass[aps,prl,showpacs,groupedaddress,twocolumn,preprintnumbers,amsmath,amssymb]{revtex4}

\usepackage{graphicx}
\usepackage{dcolumn}
\usepackage{bm}
\begin{document}

\title{Control of spin current in a Bose gas by bang-bang pulses}

\author{Yujiro Eto$^{1}$}
\author{Mark Sadgrove$^{2}$}
\author{Sho Hasegawa$^{1}$}
\author{Hiroki Saito$^{3}$}
\author{Takuya Hirano$^{1}$}
\affiliation{%
$^{1}$Department of Physics, Gakushuin University, Toshima, Tokyo 171-8588, Japan\\
$^{2}$Center for Photonic Innovations, University of Electro-Communications, Chofu, Tokyo 182-8588, Japan \\
$^{3}$Department of Engineering Science, University of Electro-Communications, Chofu, Tokyo 182-8585, Japan}

\date{\today}
             
\begin{abstract}
We generate spin currents in an $^{87}$Rb spin-2 Bose-Einstein condensate by application of a magnetic field gradient. 
The spin current destroys the spin polarization, leading to a sudden onset of two-body collisions. 
In addition, the spin coherence, as measured by the fringe contrast using Ramsey interferometry, is reduced drastically but experiences a weak revival due to in-trap oscillations. 
The spin current can be controlled using periodic $\pi$ pulses (bang-bang control), producing longer spin coherence times.
Our results show that spin coherence can be maintained even in the presence of spin currents, with applications to quantum sensing in noisy environments.
\end{abstract}

\maketitle
Understanding and manipulating spin dynamics is a major area of interest in condensed matter physics. 
In particular, the control of spin currents is a key ingredient in applications such as spintronics which has been extensively studied in solid state systems \cite{Zutic04,Fert08}. 
Although such studies originated in the field of condensed matter, related phenomena including spin segregation \cite{Lewandowski02,Du08,Du09} and the spin hole effect  \cite{Beeler13} have been demonstrated in ultracold atomic gases.
In addition, the control of spin current in quantum gases has been utilized to study superfluidity, and a rich variety of phenomena such as dark and bright solitons and modulational instability have been observed \cite{Hamner11, Hoefer11}.

A different perspective on these spin transport effects is found in the  field of quantum sensing. 
Here, spin currents can be a source of dephasing, and the associated decay of coherence and therefore sensitivity requires that spin currents be controlled.  
Dynamical decoupling techniques such as quantum bang-bang control  \cite{Viola98}, in which undesired environmental interactions are averaged out by repeated application of suitable pulses, provides one powerful solution for this purpose. 
By application of dynamical decoupling, it has been theoretically shown that spin coherence can be prolonged in a spin-1 Bose gas \cite{Ning11}. 
Furthermore, the effects of weak magnetic dipole-dipole interactions (MDDIs) have been revealed \cite{Yasunaga08} and controlled using dynamical decoupling techniques \cite{Ning12}.
Experimentally, dynamical stabilization was demonstrated in a strongly interacting spin system by application of periodic microwave pulses \cite{Hoangs13}. 

In the present study, we demonstrated the application of the bang-bang control technique to spin currents in an $^{87}$Rb $F = 2$  Bose-Einstein condensate (BEC).
We prepared a BEC in a transversely polarized spin state by applying a radio frequency (rf) pulse to the $F = 2,$ $m_F = -2$ hyperfine state.
Spin currents were then generated by application of a magnetic field gradient along the trap axis of the BEC.
We observed the spin current by measuring the relative center of mass (COM) of each spin component, providing a direct measurement of the spin transport, as opposed to mere observation of ``spin texture" in the BEC density distribution.
We observed the sudden onset of spin-changing and inelastic collisions.
In addition, we probed the spin coherence properties of the condensate in the presence of the spin current by performing Ramsey interferometry.
The interferometer fringe visibility was found to reduce drastically after a few tens of msec, with a weak revival due to in-trap oscillations of the spin current.
We show that these spin-current induced phenomena can be largely suppressed by bang-bang control consisting of periodically applied $\pi$ pulses.

We note that in Ref. \cite{Eto13PRA}, the variation of spin direction induced by an AC magnetic field at a specific frequency was extracted by using the spin echo technique with a single $\pi$ pulse which removed the effect of the slowly varying magnetic field and the helicity of the spin caused by the magnetic field gradient \cite{Eto13APEX}.  
Our results in the present paper imply that sensitive AC magnetometry is possible  \emph{even in the presence of a DC spatially inhomogeneous magnetic fields} if the bang-bang control technique is utilized. 
In addition, the application of bang-bang pulse enables control over the bandwidth of AC magnetometers  \cite{Taylor08, Grinolds13}.

We now move to the setup used in our experiments.
We produce an $^{87}$Rb BEC containing $2.8(2)\times10^5$ atoms in the hyperfine state $F = 2, m_{F} = -2 $ in a crossed far off-resonant optical dipole trap (FORT) with axial ($z$ direction) and radial frequencies of $\omega_a / (2\pi) = 20$ Hz and $\omega_r / (2\pi) = 100$ Hz (see Ref. \cite{Eto13APEX} for a more detailed description).
The bias magnetic field and its gradient along the $z$ direction (i.e. along the trap axis) are $B_{z} = 92.6$ mG and $d B_{z}/dz = 15$ mG/cm.

\begin{figure*}[t]
\includegraphics[width=16cm]{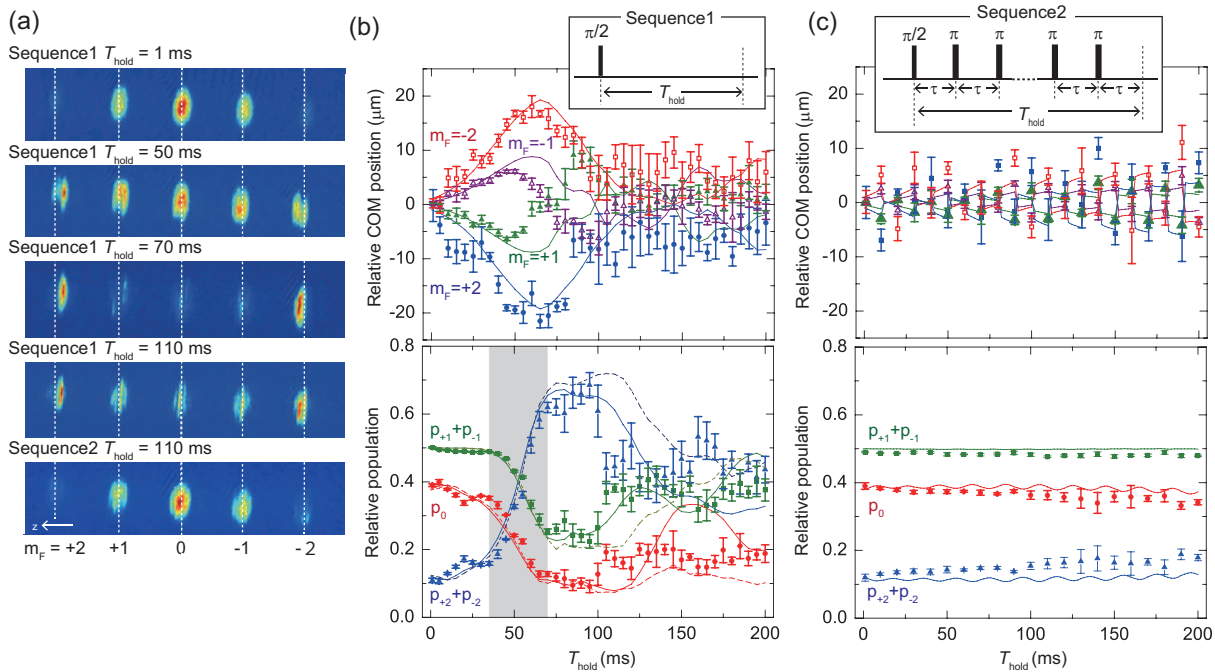}
\caption{(a) Typical absorption images of condensates obtained using sequence1 and 2. 
(b) $T_{\mathrm{hold}}$ dependence of the relative COM position of $m_{F} = \pm2$ and $\pm1$ (top) and $m_{F}$ populations (bottom).
The COM position of $m_{F} = 0$ at each $T_{\mathrm{hold}}$ is assumed to be $z = 0$ and the distance traveled by each component during the measurement is corrected using data at  $T_{\mathrm{hold}} \sim 0$.
The inset of the top panel shows the applied rf pulse sequence (sequence1).
(c) The experimental results obtained using sequence2.
$\tau$ is fixed to 10 ms. 
The inset above the top panel shows the applied rf pulse sequence (sequence2).
The solid curves in (b) and (c) indicate numerical simulation results from coupled GP equations including MDDI.
The dashed curves in bottom panel of (b) indicate numerical simulations without MDDI. 
See Supplemental Material \cite{supp}.
}
\label{fig1}
\end{figure*}

In order to observe and control the spin dynamics,
various rf pulse sequences are applied to the BEC in its initial $F = 2, m_{F} = -2$ state.
Firstly, a transversely polarized spin state is prepared by application of a $\pi/2$ pulse at the start of each of sequence inducing Larmor precession in the $x-y$ plane perpendicular to the external magnetic field,
and then the gradient magnetic field of $15$ mG/cm along $z$ direction creates the spin current.
We note that for a smaller gradient of $3$ mG/cm and similar experimental conditions, the MDDI influences the spin dynamics \cite{Eto14PRL} making this lower gradient regime less suitable for studying pure spin current effects. 
After variable holding time, $T_{\mathrm{hold}}$, the BEC is released from the FORT.
Each $m_F$ component is spatially separated along the $z$ direction using the Stern-Gerlach (SG) method.
The atomic distribution of each $m_{F}$ component is measured using absorption imaging, after the time-of-flight (TOF) of 15 ms.


Figures \ref{fig1}(a) show the typical absorption images obtained using sequence1 and 2, and each sequence is depicted  in the insets of the top row of Figs. \ref{fig1}(b) and \ref{fig1}(c) respectively.
Just after the first $\pi$/2 pulse, each $m_F$ component is populated in the ratio of $p_{0}:p_{\pm1}:p_{\pm2} = 3/8:1/4:1/16$, as shown in top image of Fig. \ref{fig1}(a),
where $p_{m_{F}}$ is the relative population for $m_{F}$ component.
As seen in Fig. 1, in the case of sequence1, hold times of a few tens of msec lead to distinct changes in the $m_F$ populations relative to the initial distribution.

In order to confirm the generation of spin current we focus on the relative position of each $m_{F}$ component.
The top panel of Fig. \ref{fig1}(b) shows the relative center of mass (COM) positions along the $z$ direction  for the $m_F=\pm1$ and $\pm2$ components,
where the COM position of the $m_F=0$ component is assumed to be zero.
When $T_{\mathrm{hold}} \lesssim 50$ ms, the positive $m_{F}$ components move  in the opposite direction to the negative $m_F$ components, 
and the distances moved by the $m_F=\pm 2$ components are about twice as large as those seen in the case of $m_F=\pm 1$.
This behavior means that the spin current is successfully induced by the spin dependent force arising from the magnetic field gradient. 
This force is proportional to  $-m_F{\rm d}B_z/{\rm d}z$.

The other feature of the observed spinor dynamics is the change of the relative population.
The bottom panel of Fig. \ref{fig1}(b) shows the spin populations ($p_{+2}+p_{-2}, p_{+1}+p_{-1}, p_{0}$) versus $T_{\mathrm{hold}}$,
where $p_{m_{F}} = N_{m_{F}}/\Sigma_{m_{F}} N_{m_{F}}$ and $N_{m_{F}}$ is the atom number for $m_{F}$ component.
The population in each $m_F$ component changes drastically after $T_{\rm hold}\sim 35$ ms [Grey area in the bottom panel of Fig. \ref{fig1}(b)].
Similar observations were observed in spin-1 and spin-2 BEC systems and so-called spin mixing \cite{Schmaliohann04, Chang04, Kuwamoto04, Chang05} where they were induced by spin-changing elastic collisions.
Spin-changing collisions do not occur for the initial state of our system produced by the application of $\pi/2$ pulse applied to the fully polarized state ($F = 2, m_F = -2$).
In the time evolution, however each $m_{F}$ component acquires a momentum $\propto m_F$ induced by the magnetic field gradient force and acquires a kinetic energy $\propto m_F ^2$. 
This energy difference creates a different phase-shift for each $m_F$ component, 
and thus the transverse spin polarization is destroyed by the spin current, which would cause the spin-changing collisions.
A similar $m_F$ dependent phase-shift can be induced by the quadratic Zeeman effect  \cite{Kronjager05,Kronjager06}, although we note that the quadratic Zeeman shift is negligible in the current experiment because the bias magnetic field is small.

\begin{figure}[t]
\includegraphics[width=7.5cm]{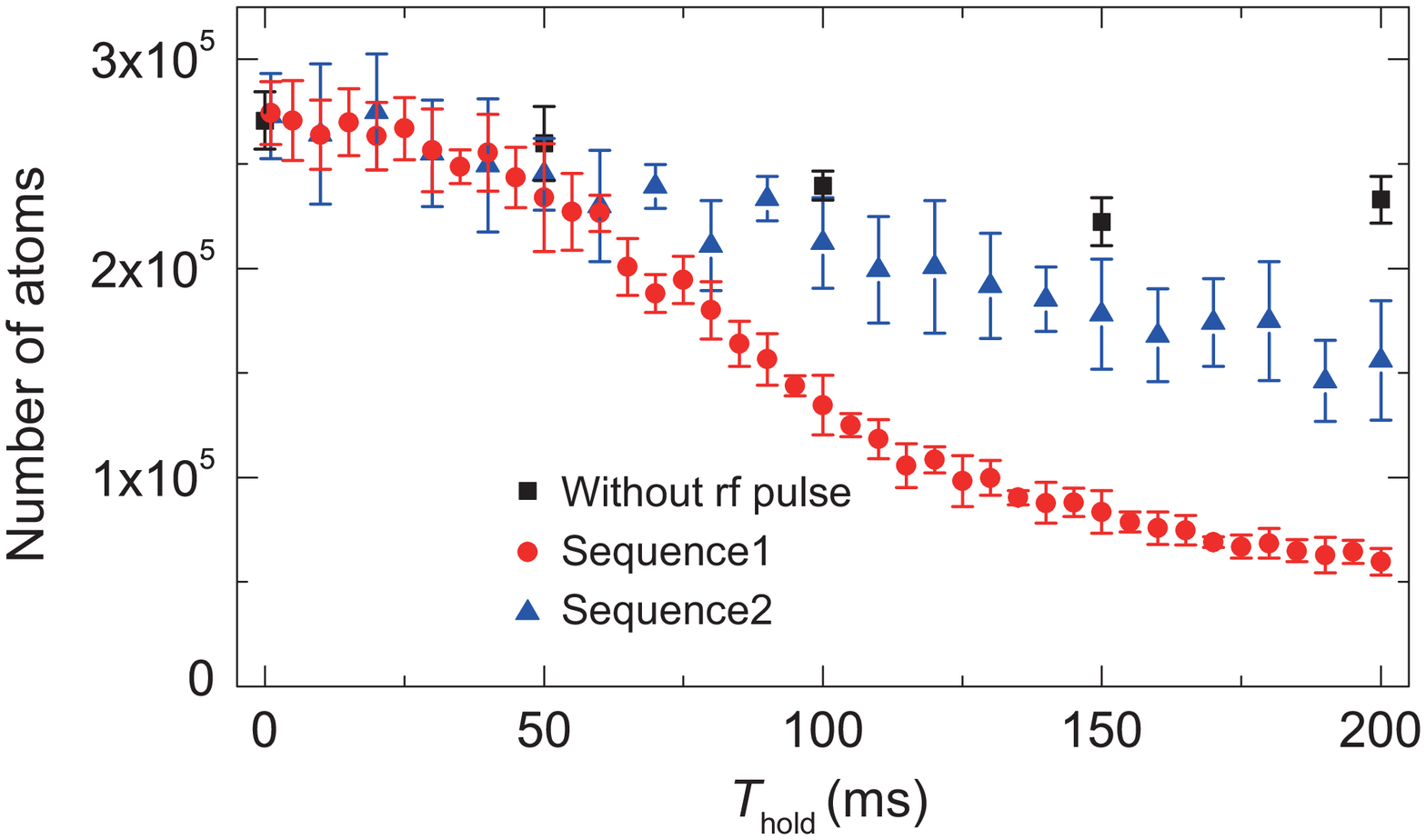}
\caption{Total atom number versus $T_{\mathrm{hold}}$.
}  
\label{fig2}
\end{figure}

The solid and dashed curves in bottom panel of Fig. \ref{fig1}(b) indicate numerical simulation results from coupled Gross-Pitaevskii (GP) equations with and without MDDIs.
Details of the simulation method are described elsewhere \cite{Eto14PRL}.
Although both curves reproduce the population oscillation, quantitative differences occur after $T_{\rm hold} \sim 75$ ms, with better agreement found in the case of MDDI inclusive simulations.
The differences seen between the simulated population dynamics in the absence and presence of MDDI are surprising given the weak nature of the MDDI in $^{87}$Rb.  
The exact role of the MDDI in the present experiment is still unclear and is currently under investigation.

Figure \ref{fig2} shows the total atom number versus $T_{\mathrm{hold}}$.
Compared with the case in which no rf pulse is applied to the BEC (the square marks in Fig. \ref{fig2}), 
the total atom number for sequence1 (the circle marks in Fig. \ref{fig2}) is drastically reduced with essentially the same timing as the precipitous change of populations seen in Fig. \ref{fig1}(b).
This enhancement of the atom loss can be understood as the increase of the hyperfine-changing inelastic collision, which is caused by the same mechanism as the spin-changing elastic collisions.

\begin{figure}[t]
\includegraphics[width=7.5cm]{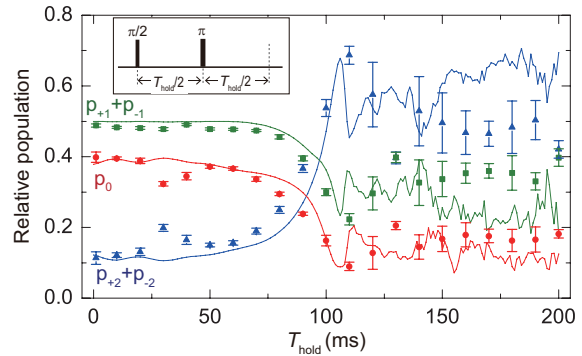}
\caption{Dynamics of the $m_F$ sublevel populations after application of a single $\pi$ pulse. The inset shows the sequence used for these measurements. The $\pi$ pulse is applied at $t = T_{\rm hold}/2$.
The solid curves indicate numerical simulation results from coupled GP equations.}
\label{fig3}
\end{figure}

In order to control the spin current and spin current-induced phenomena,
we apply a $\pi$ pulse in the middle of $T_{\mathrm{hold}}$ as shown in the inset of Fig. \ref{fig3}.
Since the $\pi$ pulse inverts the sign of $m_{F}$, the direction of the spin dependent force proportional to $-m_{F}dB_{z}/dz$ is also inverted.
We therefore expect that the spin current will be decelerated and the enhancement of spin-changing collisions should be reduced.
As shown in Fig. \ref{fig3},  
although the $m_{F}$ populations are almost constant up to $T_{\mathrm{hold}} \sim 70$ ms, 
they change drastically after 70 ms.
This population change can be understood due to the spin-changing elastic collisions induced by the spin current.
However, we note that the time of onset of the spin changing collisions has been effectively doubled from $\sim 35$ ms as seen in Fig. 1(b) to $70$ ms here.
This is due to the application of the $\pi$ pulse. 
This suggests that in order to suppress the effect of spin-changing collisions, it is important that the first $\pi$ pulse is applied before significant collisions occur.


In order to suppress the spin current induced collisions for a longer time,
we apply multiple $\pi$ pulses at constant intervals of $\tau = 10$ ms.
The bottom panel of Fig. \ref{fig1}(a) shows absorption images of the $m_F$ components of the condensate at $T_{\mathrm{hold}} =110$ ms observed using sequence2.
The observed image is seen to be almost the same as that of the initial state [top image of Fig. \ref{fig1}(a)].
The COM positions and relative populations for sequence2 are shown in Fig. \ref{fig1}(c).
The COM positions are seen to move around $z = 0$ with the relative populations of each $m_F$ level  remaining almost unchanged with increasing $T_{\rm hold}$.
Furthermore, the atom number loss for sequence2 (triangle marks in Fig. \ref{fig2}) is less than that for sequence1.
These results indicate that the spin current is suppressed by periodic application of $\pi$ pulses.
As a result, spin-changing elastic collisions and hyperfine-changing inelastic collisions are also suppressed.

\begin{figure}[t]
\includegraphics[width=7.5cm]{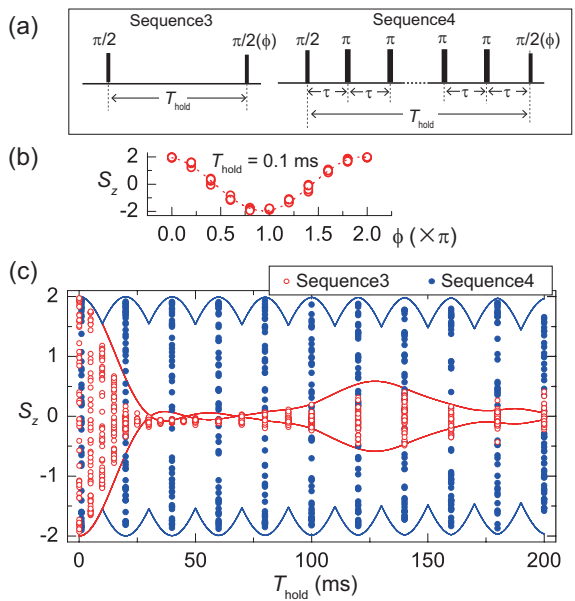}
\caption{Larmor precession signal obtained using the Ramsey interferometer.
(a) Sequences used to realize the Ramsey interferometer, where $\tau = 10$ ms.  
(b) $S_{z}$ versus $\phi$ at $T_{\mathrm{hold}} = 0.1$ ms. 
(c) $S_{z}$ versus $T_{\mathrm{hold}}$.
The solid curves indicate envelopes of numerical simulation results from coupled GP equations.
}
\label{fig4}
\end{figure}

We have seen that the application of multiple $\pi$ pulses with an appropriate repetition period can control the spin current and spin current induced phenomena.
We now assess the ability of the bang-bang control method to preserve coherence by using a Ramsey interferometry sequence. 
Firstly, absorption images are acquired for various relative phases $\phi$ between the two $\pi/2$ pulses in sequence3 [left sequence of Fig. \ref{fig4}(a)] to probe the Larmor precession in the $x-y$ plane \cite{Mark13}. 
From the obtained images, we calculate the expectation value of $z$-component of the spin,
$S_{z} = (\sum_{m_{F}} m_{F} N_{m_{F}})/(\sum_{m{F}} N_{m_{F}})$.
Figure \ref{fig4}(b) shows $S_{z}$, versus $\phi$ at $T_{\mathrm{hold}} = 0.1$ ms.
The essentially perfect contrast of $S_z$ seen is expected for the very short hold time of $T_{\rm hold}=0.1$ ms used here.

In order to observe the long time behavior,
$S_{z}$ is measured for various values of $T_{\mathrm{hold}}$.
Figure \ref{fig4}(c) shows the values of $S_{z}$ as a function of $T_{\mathrm{hold}}$. 
In the case where sequence3 is used [open circles in Fig. \ref{fig4}(c)],
the signal contrast of the Ramsey interferometer is drastically reduced up to $T_{\mathrm{hold}} \sim 20$ ms.

In our definition of $S_{z}$, the spin density is averaged over the whole spatial region.
The contrast of $S_{z}$ is thus largely reduced due to the spatial variations of spin direction, that is, the magnetic field gradient-induced spin helicity.
For a magnetic field gradient of $dB_z/dz = 15$ mG/cm, 
the distance $\Delta z$ along which the spin helix is twisted by $\pi$ for $T = 20$ ms is estimated to be $\pi\hbar/[g_{F} \mu_B (dB_z/dz) T] \simeq 24$ $\mu{\rm m}$ with $g_{F}$ and $\mu_B$ being the g-factor and the Bohr magneton, 
which is comparable to the Thomas-Fermi radius in the $z$ direction $\simeq 29$ $\mu{\rm m}$.

After the rapid reduction, a slight revival of the contrast is seen for $T_{\rm hold} >100$ ms. 
This revival is clearly reproduced in the simulation [solid curves in Fig. \ref{fig4}(c)].
This behavior implies that the spin polarization is partially realigned as a result of the oscillation of the spin current in the harmonic optical trapping potential along with the spin-changing collisions.
It is interesting to note that such a revival was \emph{not} observed in the case of a $F = 1$ elongated BEC~\cite{Higbie05}.

To prolong the coherence time of the Larmor precession, we applied multiple $\pi$ pulses at constant intervals of $\tau = 10$ ms in order to remove spin-current induced effects.
The filled circles in Fig. \ref{fig4}(c) show $S_{z}$ obtained using sequence4.
The signal contrast is almost unchanged up to  $T_{\mathrm{hold}} = 200$ ms due to the application of the $\pi$ pulses.
This result demonstrates that bang-bang control is a very effective method for prolonging the spin coherence time.


In conclusion, we have investigated the effects of spin current in an $^{87}$Rb $F=2$ Bose-Einstein condensate.
A transversely polarized initial spin state is prepared which evolves in the presence of the magnetic field gradient of 15 mG/cm. 
A gradient induced spin current and subsequent spin-changing collisions were observed.
In addition, we probed the Larmor precession of the spin using a Ramsey interferometer.
It was found that the contrast of the precession signal is drastically reduced after a few tens of msec. 
We showed that these various effects can be greatly suppressed by controlling the spin current. 
The technique shown in this paper is thus useful for maintaining coherence in quantum sensing applications, with a particular example being magnetometry using a BEC.
In particular, we note that sequence4 which consists of a Ramsey interferometer with multiple $\pi$ pulses is directly applicable to techniques  of AC magnetometry \cite{Grinolds13} and spectrum analyzers of the environment \cite{Almong11, Kotler13}.
Lastly, we note that while the present work has focused on suppressing the effects of the spin current, in future experiments it may be possible to use the techniques considered here to enhance or otherwise control the spin current rather than suppressing it, opening the door to more fundamental studies of spin transport in spinor BECs.

This work was supported by the Japan Society for the Promotion of Science (JSPS) through its Funding Program for World-Leading Innovation R\&D on Science and Technology (FIRST Program), 
and a Grant-in-Aid for Scientific Research (C) (No. 23540464 and No. 26400414) and a Grant-in-Aid for Scientific Research on Innovation Areas Fluctuation \& Structure (No. 25103007) from the Ministry of Education, Culture, Sports, Science, and Technology of Japan.


\begin{thebibliography}{99}
\bibitem{Zutic04}
I. \^Zuti\'c, J. Fabian, and S. D. Sarma, 
Rev. Mod. Phys. {\bf 76,} 323 (2004).

\bibitem{Fert08}
A. Fert, 
Rev. Mod. Phys. {\bf 80,} 1517 (2008).

\bibitem{Lewandowski02}
H. J. Lewandowski, D. M. Harber, D. L. Whitaker, and E. A. Cornell,
Phys. Rev. Lett. {\bf 88,} 070403 (2002).

\bibitem{Du08}
X. Du, L. Luo, B. Clancy, and J. E. Thomas,
Phys. Rev. Lett. {\bf 101,} 150401 (2008).

\bibitem{Du09}
X. Du, Y. Zhang, J. Petricka, and J. E. Thomas,
Phys. Rev. Lett. {\bf 103,} 010401 (2009).

\bibitem{Beeler13}
M. C. Beeler, R. A. Williams, K. Jim\'enez-Garc\'ia, L. J. LeBlanc, A. R. Perry, and I. B. Spielman,
Nature {\bf 498,} 201 (2013).

\bibitem{Hamner11}
C. Hamner, J. J. Chang, P. Engels, and M. A. Hoefer, 
Phys. Rev. Lett. {\bf 106,} 065302 (2011).

\bibitem{Hoefer11}
M. A. Hoefer, J. J. Chang, C. Hamner, and P. Engels,
Phys. Rev. A {\bf 84,} 041605(R) (2011).

\bibitem{Viola98}
L. Viola and S. Lloyd,
Phys. Rev. A {\bf 58,} 2733 (1998).

\bibitem{Ning11}
B.-Y. Ning, J. Zhuang, J. Q. You, and W. Zhang,
Phys. Rev. A {\bf 84,} 013606 (2011).

\bibitem{Yasunaga08}
M. Yasunaga and M. Tsubota,
Phys. Rev. Lett. {\bf 101,}  220401 (2008).

\bibitem{Ning12}
B.-Y. Ning, S. Yi, J. Zhuang, J. Q. You, and W. Zhang,
Phys. Rev. A {\bf 85,} 053646 (2012).

\bibitem{Hoangs13}
T. M. Hoang, C. S. Gerving, B. J. Land, M. Anquez, C. D. Hamley, and M. S. Chapman,
Phys. Rev. Lett. 111 090403 (2013). 

\bibitem{Eto13PRA}
Y. Eto, S. Sekine, S. Hasegawa, M. Sadgrove, H. Saito and T. Hirano,
Appl. Phys. Express {\bf 6,} 052801 (2013).

\bibitem{Eto13APEX}
Y. Eto, H. Ikeda, H. Suzuki, S. Hasegawa, Y. Tomiyama, S. Sekine, M. Sadgrove, and T. Hirano,
Phys. Rev. A {\bf 88,} 031602(R) (2013). 

\bibitem{Taylor08}
J. M. Taylor, P. Cappellaro, L. Childress, L. Jiang, D. Budker, P. R. Hemmer, A. Yacoby, R. Walsworth, and  M. D. Lukin,
Nature Physics {\bf 4,} 810 (2008). 

\bibitem{Grinolds13}
M. S. Grinolds, S. Hong, P. Maletinsky, L. Luan, M. D. Lukin, R. L. Walsworth, and A. Yacoby,
Nature Phys. {\bf 7,} 320 (2013). 

\bibitem{supp}
See Supplemental Materials for movies showing time evolution of $m_{F}$ components for sequence1 and 2 obtained by numerical simulations.

\bibitem{Eto14PRL}
Y. Eto, H. Saito, T. Hirano,
to appear in Phys. Rev. Lett.

\bibitem{Schmaliohann04}
H. Schmaljohann, M. Erhard, J. Kronj\"{a}ger, M. Kottke, S. van Staa, L. Cacciapuoti, J. J. Arlt, K. Bongs, and K. Sengstock,
Phys. Rev. Lett. {\bf 92,} 040402 (2004).

\bibitem{Chang04}
M.-S. Chang, C. D. Hamley, M. D. Barrett, J. A. Sauer, K. M. Fortier, W. Zhang, L. You, and M. S. Chapman,
Phys. Rev. Lett. {\bf 92,} 140403 (2004).

\bibitem{Kuwamoto04}
T. Kuwamoto, K. Araki, T. Eno, and T. Hirano,
Phys. Rev. A {\bf 69,} 063604 (2004).

\bibitem{Chang05}
M.-S. Chang, Q. Qin, W. Zhang, L. You, and M. S. Chapman,
Nature Phys. {\bf 1,} 111 (2005).

\bibitem{Kronjager05}
J. Kronj\"ager, C. Becker, M. Brinkmann, R. Walser, P. Navez, K. Bongs, and K. Sengstock, 
Phys. Rev. A {\bf 72,} 063619 (2005).

\bibitem{Kronjager06}
J. Kronj\"ager, C. Becker, P. Navez, K. Bongs, and K. Sengstock, 
Phys. Rev. Lett {\bf 97,} 110404 (2006).

\bibitem{Mark13}
M. Sadgrove, Y. Eto, S. Sekine, H. Suzuki and T. Hirano,
J. Phys. Soc. Jpn. {\bf 82,} 094002 (2013). 

\bibitem{Higbie05}
J. M. Higbie, L. E. Sadler, S. Inouye, A. P. Chikkatur, S. R. Leslie, K. L. Moore, V. Savalli, and D. M. Stamper-Kurn,
Phys. Rev. Lett. {\bf 95,} 050401 (2005).

\bibitem{Almong11}
I. Almog, Y. Sagi, G. Gordon, G. Bensky, G. Kurizki, and N. Davidson, 
J. Phys. B {\bf 44,} 154006 (2011).

\bibitem{Kotler13}
S. Kotler, N. Akerman, Y. Glickman, and R. Ozeri,
Phys. Rev. Lett. {\bf 110,} 110503 (2013).



\end{thebibliography}
\end{document}